\documentclass{article}

\PassOptionsToPackage{square,numbers}{natbib}

    \usepackage[preprint]{neurips_2020}

\usepackage[utf8]{inputenc} 
\usepackage[T1]{fontenc}    
\usepackage{hyperref}       
\usepackage{url}            
\usepackage{booktabs}       
\usepackage{amsfonts}       
\usepackage{nicefrac}       
\usepackage{microtype}      
\usepackage{graphicx}
\usepackage{subfig}
\usepackage{caption}
\title{Data Augmentation for Electrocardiogram Classification with Deep Neural Network}

\author{%
  Naoki Nonaka \\
  Medical Innovations Hub Program\\
  Riken\\
  Tokyo, Japan \\
  \texttt{naoki.nonaka@riken.jp} \\
  \And
  Jun Seita \\
  Medical Innovations Hub Program\\
  Riken\\
  Tokyo, Japan \\
  \texttt{jun.seita@riken.jp} \\
}

\begin{document}

\maketitle

\begin{abstract}
Electrocardiogram (ECG) is the most crucial monitoring modality to diagnose cardiovascular events.
Precise and automatic detection of abnormal ECG patterns is beneficial to both physicians and patients.
In the automatic detection of abnormal ECG patterns, deep neural networks (DNNs) have shown significant achievements.
However, DNNs require large amount of labeled data, which are often expensive to obtain.
On the other hand, recent research have shown by randomly combining data augmentations can improve image classification accuracy.
Thus, in this work we explore data augmentation suitable for ECG data and propose ECG Augment.
We show by introducing ECG Augment, we can improve classification of atrial fibrillation with single lead ECG data, without changing an architecture of DNN.

\end{abstract}

\section{Introduction}

Electrocardiogram (ECG) is widely used device to monitor hearts' electronic activities.
This monitoring is crucial to diagnose cardiovascular diseases.
Atrial fibrillation is one of the heart's abnormal activities which are associated with stroke, heart failure, and death.
In practice, to diagnose abnormal heart activities, cardiologist review ECG signals which requires labor-intensive process putting large burden on cardiologist.
Thus, to alleviate this labor intensive processes, an automatic detection system of abnormal ECG to assist physicians are developed.

In order to automatically detect abnormal ECGs, various machine learning approaches have been proposes.
Among these approaches, several works have shown deep neural networks (DNNs) can detect irregular ECGs without the needs of feature engineering \citep{hannun2019cardiologist,attia2019artificial}.
DNNs have driven substantial advances and demonstrated dramatic improvement of state of the art in tasks like image recognition, machine translation and speech recognition \citep{szegedy2015going,he2016deep,bahdanau2014neural,devlin2018bert,graves2013speech}.
However, DNNs often requires large amount of labeled training data in to achieve good performance.
In the context of medical data, it is often costly to acquire large amount of labeled data, which requires intensive work of medical expert.

Data augmentation is a technique for improving classification accuracy used in domains such as image classification.
Data augmentation increases amount and diversity of data by adding random perturbation based on augmentation strategy.
For instance, in image domain, augmentations includes flipping image or adding some pixels.
Recent work have shown classification accuracies can be improved by randomly combining multiple augmentations \citep{cubuk2018autoaugment,cubuk2020randaugment}.

Thus, in this work we explored data augmentation technique suitable for ECG data and propose ECG Augment.
We show by introducing ECG Augment, we can improve classification of atrial fibrillation with single lead ECG data, without changing an architecture of DNN.

\section{Related works}
In this study, we classify ECG data with DNNs.
Hence, in this section, we introduce related ECG classification methods with DNNs.

\subsection{Classification of ECG}

Various approaches which adopt hand-crafted features have long been studied, and still performs well with limited amount of data.
In CinC/Challenge 2017 \citep{clifford2017af}, top ranked teams extract more than 50 features and trained model with extracted features to obtain final model \citep{datta2017identifying,hong2017encase,teijeiro2017arrhythmia,zabihi2017detection,mahajan2017cardiac}.
Approaches with hand-crafted features have shown to be effective with limited amount of data.
However, these approaches usually requires time consuming feature engineering and processing which relies on domain knowledge and experiences.

Wide variety of research fields have experienced advances with DNNs, including the field of ECG classification, with its ability to automatically extract features.
Several researches with end-to-end DNN applied to ECG classification related tasks have been conducted.
\citet{hannun2019cardiologist} have shown a cardiologist level classification accuracy with there data using a 34 layers DNN. 
Also, \citet{attia2019artificial} applied DNN to 12 lead ECGs and shown the model can detect patient who previously had atrial fibliration.
In \citet{porumb2020precision}, DNN was applied to detect hypoglycemic events from ECGs.
Approaches with DNNs have shown promising results, however, deep models require large amount of data, for instance \citet{hannun2019cardiologist} collected 64,121 labeled data from 29,163 patients, \citet{attia2019artificial} collected 649,931 normal sinus rhythms from 180,922 patients and in CinC/Challenge 2017 public dataset number of samples used to train the model was 8,528.

\subsection{Data augmentation}

Data augmentation is used as a technique to improve the robustness of deep learning models.

\section{ECG Augmentation}

In this section we explain ECG augmentation.

Data augmentation is a technique to improve robustness of model by adding perturbation to input data.
Recently, it has been shown that AutoAugment \citep{cubuk2018autoaugment}, which use reinforcement learning to find the best combination of multiple data expansion methods, and RandAugment \citep{cubuk2020randaugment}, which randomly combine multiple data augmentation methods, can improve the accuracy of image classification without changing the structure of the model.
These methods improve the accuracy by combining data expansion augmentation applicable to images.
In this study, we experiment with a random combination of data augmentation methods for ECG data.

The following methods are used for data augmentation.
	
\begin{itemize}
  \item Erase: Randomly select a lead and set the signal of the selected lead to 0.
  \item Scale: Randomly scaling data.
  \item Flip: Flip the signal up and down at random.
  \item Drop: Randomly missing signal values.
  \item Cutout: Set a random interval signal to 0.
  \item Shift: Shifts the signal at random.
  \item Sine: Add a sine wave to the entire sample.
  \item Square: Add a square pulse to the entire sample.
  \item Partial sine: Add a sine wave to a random interval only.
  \item Partial square: Add a square pulse to a random interval only.
  \item Partial white noise: Add white noise to a random interval.
  \item FIR low: A finite impulse response filter is applied as a low pass filter.
  \item FIR high: A finite impulse response filter is applied as a high pass filter.
\end{itemize}

The additional hyper parameters of augmentation is magnitude of each augmentation, $M$ and number of augmentation to apply in each batch, $N$.

\section{Experiment}

In this section we present the training settings of our experiment.

\subsection{Dataset}

In this section we describe the details of dataset used in the experiment.

\subsubsection{CinC/Challenge 2017 dataset}

The CinC/Challenge 2017 dataset consists of ECG data collected by AliveCor device donated by AliveCor.
Each sample is 9 to 61 seconds long, collected at a frequency of 300 Hz, a range of $\pm{5}$ mV, and a bandwidth of 0.5 - 40 Hz.
We conducted an experiment using the public dataset of the CinC/Challenge 2017 dataset, which consist of 8,528 samples.
Each sample in the data set is assigned one of four class labels: Normal, atrial fibliration (AF), other rhythm, and too noisy.
The number of class labels included in the data set is 5154 for normal rhythm, 771 for AF rhythm, 2557 for other rhythm, and 46 for noisy.

We applied three processing to original CinC/Challenge 2017 data, scaling, padding and down sampling.
We first scaled the data by dividing the raw ECG signals by absolute value of dynamic range value, namely 500.
Consecutively we applied down sampling to scaled ECG data.
Original CinC/Challenge 2017 collected at a frequency of 50 Hz with data length ranging from 9 seconds to 61 seconds.
Finally, in order to align the length of samples, we added zero vector to head of each samples until length of vector aligns with the longest sample contained in dataset.
As a result of processing procedure, all the samples contained in CinC/Challenge 2017 dataset have length of 3050 dimension.

\subsection{Evaluation criteria}

Following the evaluation criteria for CinC/Challenge 2017 \citep{clifford2017af}, we calculate F1 score for 3 classes (Normal, AF and Other rhythms).
We denote F1score of the model for class $c$ by $F1score_c$.
The final score of the model was calculated by averaging F1score of three classes.
\begin{math}
score = \frac{F1score_{Normal} + F1score_{AF} + F1score_{other}}{3}
\end{math}

The CinC/Challenge 2017 dataset was split in to train, validation and test dataset with size ratio of 0.6, 0.2 and 0.2 respectively.
We conducted 5 independent trials for each training settings.
For each trials and training settings, samples in train, validation and test dataset was randomly shuffled.

\subsection{Model architecture}

\begin{figure*}[ht]
  \subfloat[Overall architecture]{
	\begin{minipage}[t]{0.65\textwidth}
	   \centering
	   \includegraphics[width=1\textwidth]{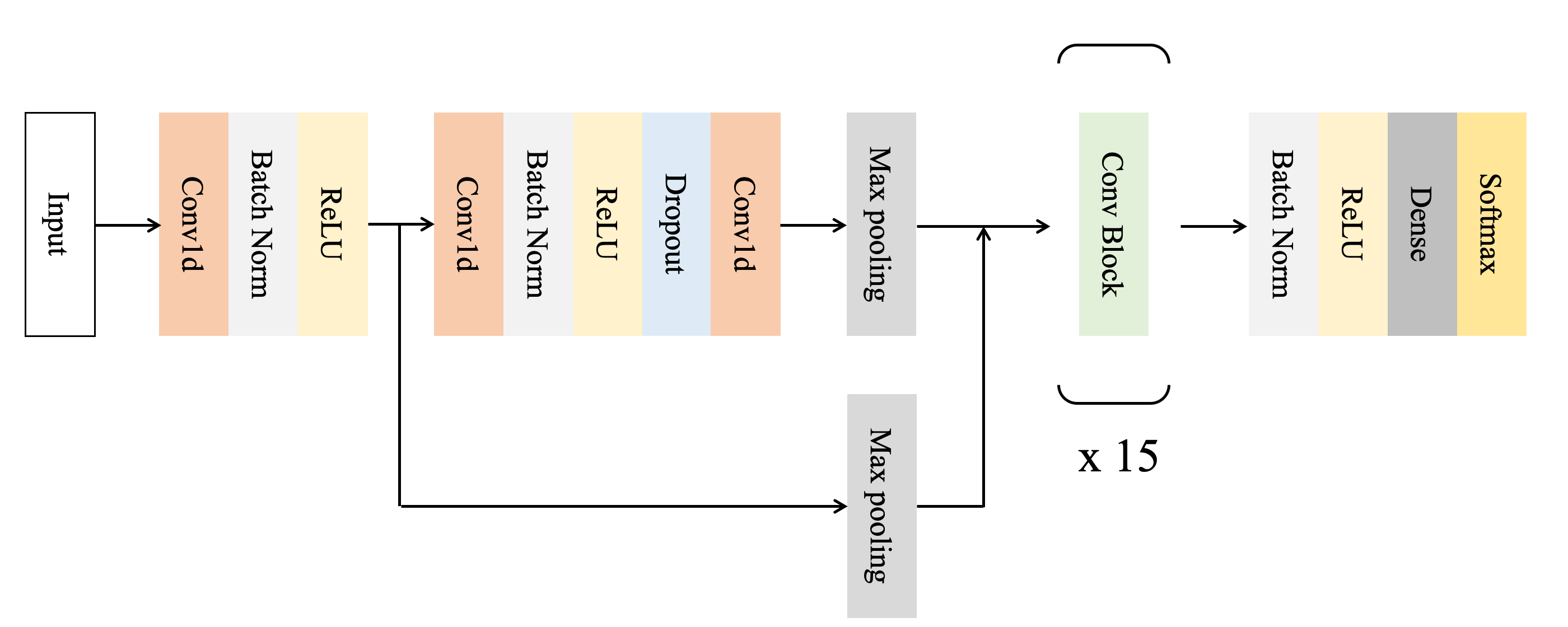}
       \label{fig:model_a}
    \end{minipage}} 
  \subfloat[Conv Block]{
	\begin{minipage}[t]{0.3\textwidth}
	   \centering
	   \includegraphics[width=1\textwidth]{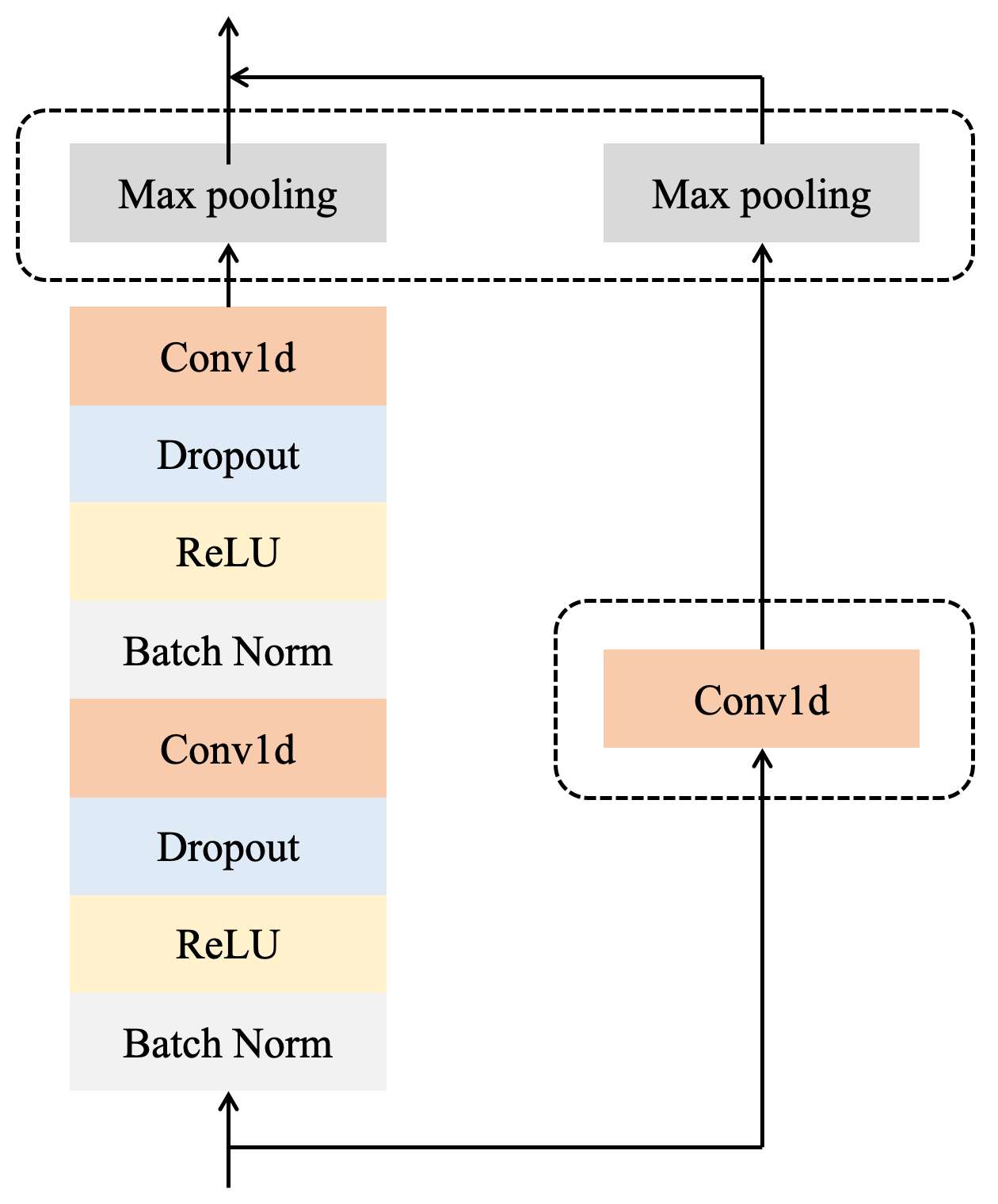}
       \label{fig:model_b}
	\end{minipage} } 
\caption{Model architecture of ECG classification model. Figure (a) shows overall architecture and Conv Block shown in Figure (b) is stacked 15 times in the model.}
\label{fig:model}
\end{figure*}

In this section, we explain the model architecture of DNN used for classification of ECG data.
We adopt a convolutional neural network for the ECG sequence classification task. 
Similar to the work of  \citet{andreotti2017comparing}, \citet{hannun2019cardiologist} and \citet{li2019domain}, our model employ shortcut connections which is similar to Residual Network \citep{he2016deep} to allow information to propagate well in very DNNs.
Figure \ref{fig:model_a} shows a high-level architecture of the network used for ECG classification.
The network takes a time-series of raw ECG signal as an input, and outputs a label predictions for each sequences.
Our model consist of 15 ConvBlocks, shown as in Figure \ref{fig:model_b}, and followed by a fully connected layer and a softmax activation.
Each ConvBlocks has 2 convolutional layers with filter length of 16 and have 8k filters, where k is 1 at start and incremented every 2 ConvBlocks.
Max pooling layer in ConvBlock is applied in every 2 blocks, beginning from first ConvBlock.

We apply both Batch Normalization \citep{ioffe2015batch} and Dropout \citep{srivastava2014dropout} before convolution in each ConvBlock, and insert rectified linear unit, ReLU after every Batch Normalization layer. 
The final fully connected layer and softmax activation outputs vector of length 4, which corresponds to the number of classes in the classification task.

\section{Results and discussion}

In this section, we describe the results obtained from the experiment.
In this section, the case without augmentation is used as the baseline.
Augmentation has two hyperparameters for ECG augmentation: intensity and the number of augmentations used.
Experiments were performed with three different intensities and a number of augmentations of 1, 2, 3, and 5 augmentations.
The results are shown in Table \ref{tab:rand_pnet17}.
At all three intensities and the number of Augmentations at four levels of intensity and four Augmentations, the results of ECG Augmentation were above the baseline.
In particular, when the intensity of Augmentation was set to moderate and the number of Augmentations was set to 5, the mean F1score was improved by 3.17\% compared to the baseline.


\begin{table}[htbp]
 \centering 
 \caption{Average F1 score of 3 classes}
 \begin{tabular}{l|cc|cc}\hline\hline
                            & M & N & Score & Improvement (\%) \\\hline
 Baseline            & - & - & 0.8168 & - \\\hline
 ECG Augment & 4 & 1 & 0.8365 & 2.41 \\
 ECG Augment & 4 & 2 & 0.8341 & 2.12 \\
 ECG Augment & 4 & 3 & 0.8344 & 2.15 \\
 ECG Augment & 4 & 5 & 0.8342 & 2.13 \\\hline
 ECG Augment & 12 & 1 & 0.8381 & 2.61 \\
 ECG Augment & 12 & 2 & 0.8392 & 2.74 \\
 ECG Augment & 12 & 3 & 0.8418 & 3.06 \\
 ECG Augment & 12 & 5 & \bf{0.8427} & \bf{3.17} \\\hline
 ECG Augment & 20 & 1 & 0.8347 & 2.19 \\
 ECG Augment & 20 & 2 & 0.8339 & 2.09 \\
 ECG Augment & 20 & 3 & 0.8383 & 2.63 \\
 ECG Augment & 20 & 5 & 0.8287 & 1.46 \\\hline\hline
\end{tabular}
\label{tab:rand_pnet17}
\end{table}

\section{Conclusion}
In this study, a method of data augmentation in a DNN model to classify ECG data is proposed.
It is shown that the proposed method can improve the accuracy of the classification of atrial fibrillation for single induction ECGs.
The present study shows that ECG augmentation is effective in classifying atrial fibrillation.
On the other hand, ECG augmentation may not be directly applicable to arrhythmias other than atrial fibrillation, because the shape of the waveform is different.
Moreover, effective Augmentation method may be different depending on the type of induction used other than the class.
In such cases, the accuracy may be improved by using the AutoAugment method which selects an augmentation method by reinforcement learning.

\bibliographystyle{abbrvnat}

\end{document}